%% file: main.tex
\pdfoutput=1
\documentclass[conference]{IEEEtran}
 \IEEEoverridecommandlockouts
\ifCLASSOPTIONcompsoc
  \usepackage[nocofbmpress]{cite}
  \usepackage{multirow}
  \usepackage{booktabs}
  \usepackage{float}
\else
  \usepackage{cite}
\fi
\ifCLASSINFOpdf
  \usepackage[pdftex]{graphicx}
\else
\fi

\usepackage{xcolor}
\usepackage{amsmath}
\usepackage{threeparttable}
\usepackage{siunitx}
\usepackage{glossaries}
\usepackage{amssymb}
 \usepackage{booktabs} 
 \usepackage{multirow}
\usepackage{listings}
\usepackage{courier}
\usepackage[normalem]{ulem} 
                
\usepackage[hidelinks = true]{hyperref}

\renewcommand{\baselinestretch}{0.940}

\begin{document}
\bstctlcite{IEEE:BSTcontrol}

\title{VEXP: A Low-Cost RISC-V ISA Extension for Accelerated Softmax Computation in Transformers}

\author{
    \IEEEauthorblockN{Run Wang\IEEEauthorrefmark{1},  
Gamze Islamoglu\IEEEauthorrefmark{1}, Andrea Belano\IEEEauthorrefmark{2}, Viviane Potocnik\IEEEauthorrefmark{1} } 
    \IEEEauthorblockN{Francesco Conti\IEEEauthorrefmark{2}, Angelo Garofalo\IEEEauthorrefmark{2}, Luca Benini\IEEEauthorrefmark{1}\IEEEauthorrefmark{2}}
    
    \IEEEauthorblockA{\IEEEauthorrefmark{1}IIS, ETH Zurich, Switzerland \\ runwang@ethz.ch, \{gislamoglu, vivianep, lbenini\}@iis.ee.ethz.ch} 

    \IEEEauthorblockA{\IEEEauthorrefmark{2}DEI, University of Bologna, Italy \\ \{andrea.belano2, f.conti, angelo.garofalo\}@unibo.it} 

    \thanks{This work was supported by the NeuroSoC project, funded under the European Union's Horizon Europe research and innovation programme (Grant Agreement No.~101070634).}  

}

\maketitle

\input{acronyms}

\begin{abstract}
While Transformers are dominated by \gls{fp} Matrix-Multiplications, their aggressive acceleration through dedicated hardware or many-core programmable systems has shifted the performance bottleneck to non-linear functions like Softmax. Accelerating Softmax is challenging due to its non-pointwise, non-linear nature, with exponentiation as the most demanding step. To address this, we design a custom arithmetic block for Bfloat16 exponentiation leveraging a novel approximation algorithm based on Schraudolph’s method, and we integrate it into the \gls{fpu} of the RISC-V cores~\cite{zaruba_snitch_2021} of a compute cluster, through custom \gls{isa} extensions, with a negligible area overhead of \qty[detect-all=true]{1}{\percent}. By optimizing the software kernels to leverage the extension, we execute Softmax with 162.7$\times$ less latency and 74.3$\times$ less energy compared to the baseline cluster, achieving an 8.2$\times$ performance improvement and 4.1$\times$ higher energy efficiency for the FlashAttention-2 kernel in GPT-2 configuration. Moreover, the proposed approach enables a multi-cluster system to efficiently execute end-to-end inference of pre-trained Transformer models, such as GPT-2, GPT-3 and ViT, achieving up to 5.8$\times$ and 3.6$\times$ reduction in latency and energy consumption, respectively, without requiring re-training and with negligible accuracy loss.

\end{abstract}




\input{TEXT/01_Introduction_v2}
\input{TEXT/02_Related_Work}
\input{TEXT/03_Background}
\input{TEXT/04_HW_SW_Optimizations_v2}

\input{TEXT/05_Evaluation}

\input{TEXT/06_Comparison_with_SoA}
\input{TEXT/07_Conclusion}
\input{TEXT/08_Acknowledgement}



\bibliographystyle{IEEEtran}
\bibliography{references, ieeetran}

\end{document}

%% file: acronyms.tex
\newacronym[plural=CNNs, firstplural=Convolutional Neural Networks (CNNs)]{cnn}{CNN}{Convolutional Neural Network}
\newacronym[plural=DNNs, firstplural=Deep Neural Networks (DNNs)]{dnn}{DNN}{Deep Neural Network}
\newacronym[plural=LLMs, firstplural=Large Language Models (LLMs)]{llm}{LLM}{Large Language Model}
\newacronym[plural=SoCs, firstplural=System-on-Chips (SoCs)]{soc}{SoC}{System-on-Chip}
\newacronym[plural=ASICs, firstplural=Application-Specific Integrated Circuits (ASICs)]{asic}{ASIC}{Application-Specific Integrated Circuit}
\newacronym[plural=FMAs, firstplural=Fused Multiply-Add Units (FMAs)]{fma}{FMA}{Fused Multiply-Add}
\newacronym[plural=ViTs, firstplural=Vision Transformers (ViTs)]{vit}{ViT}{Vision Transformer}
\newacronym[plural=FPGAs, firstplural=Field-Programmable Gate Arrays (FPGAs)]{fpga}{FPGA}{Field-Programmable Gate Array}
\newacronym[plural=RNNs, firstplural=Recurrent Neural Networks (RNNs)]{rnn}{RNN}{Recurrent Neural Network}
\newacronym[plural=GPUs, firstplural=Graphics Processing Units (GPUs)]{gpu}{GPU}{Graphics Processing Unit}
\newacronym[plural=CPUs, firstplural=Central Processing Units (CPUs)]{cpu}{CPU}{Central Processing Unit}
\newacronym[plural=NPUs, firstplural=Neural Processing Units (NPUs)]{npu}{NPU}{Neural Processing Unit}
\newacronym[plural=ReLUs, firstplural=Rectified Linear Units (ReLUs)]{relu}{ReLU}{Rectified Linear Unit}
\newacronym{sgd}{SGD}{Stochastic Gradient Descent}
\newacronym{isa}{ISA}{Instruction Set Architecture}
\newacronym{pace}{PACE}{Polynomial Approximation Compute Engine}
\newacronym{simd}{SIMD}{Single Instruction Multiple Data}
\newacronym{rv}{RISC-V}{Reduced Instruction Set Computing - V}
\newacronym[plural=TPUs, firstplural=Tensor Processing Units (TPUs)]{tpu}{TPU}{Tensor Processing Unit}
\newacronym{ram}{RAM}{Random Access Memory}
\newacronym{lut}{LUT}{Lookup Table}
\newacronym{nlp}{NLP}{Natural Language Processing}

\newacronym[plural=CSRs, firstplural=Control Status Registers (CSRs)]{csr}{CSR}{Control Status Register}
\newacronym{ai}{AI}{Artificial Intelligence}
\newacronym{genai}{GenAI}{Generative AI}
\newacronym[plural=SSRs, firstplural=Stream Semantic Registers (SSRs)]{ssr}{SSR}{Stream Semantic Register}
\newacronym{frep}{FREP}{Floating-point Repetition}

\newacronym[plural=GEMMs, firstplural=General Matrix-Matrix Multiplications (GEMMs)]{gemm}{GEMM}{General Matrix-Matrix Multiplication}
\newacronym{bf16}{BF16}{Brain Floating-Point (Bfloat16 or BF16)}
\newacronym[plural=FPUs, firstplural=Floating-Point Units (FPUs)]{fpu}{FPU}{Floating-Point Unit}
\newacronym[firstplural=Floating-Point (FP)]{fp}{FP}{Floating-Point}
\newacronym[firstplural=Fixed-Point (FX)]{fx}{FX}{Fixed-Point}
\newacronym[firstplural=Coordinate Rotation Digital Computers (CORDICs)]{cordic}{CORDIC}{Coordinate Rotation Digital Computer}
\newacronym{spm}{SPM}{scratchpad memory}
\newacronym{dma}{DMA}{direct memory access}

%% file: TEXT/01_Introduction_v2.tex
\section{Introduction}

Transformer-based models such as the GPT family~\cite{guo_gptqt_2024} and the LLaMa family~\cite{touvron_llama_2023}, have emerged as a cornerstone of machine learning, demonstrating state-of-the-art performance in diverse domains, including natural language processing (NLP), computer vision, and audio processing. These models leverage pre-trained representations on large-scale unlabeled datasets, enabling remarkable accuracy improvements in fine-tuned downstream tasks such as sentence classification and question answering. At the core of their success is the Transformer architecture \cite{vaswani_attention_2017}, which utilizes the self-attention mechanism to model complex relationships within input sequences. 

Despite the interest in deploying Transformer-based models on mobile and edge devices, their substantial computational and memory requirements present challenges in meeting the resource and energy constraints of these devices. In encoders and the prefill stage of decoders, the computational complexity of attention layers scales quadratically with the input sequence length, leading to memory and computational overheads that necessitate mitigation by means of dedicated acceleration. Although many architectures utilize \gls{gemm} acceleration to alleviate the computational burden, performance bottlenecks are increasingly shifting toward non-linear operations, especially the Softmax function within the attention layers.

\begin{figure}[t]
\centering
\includegraphics[width=0.9\linewidth]{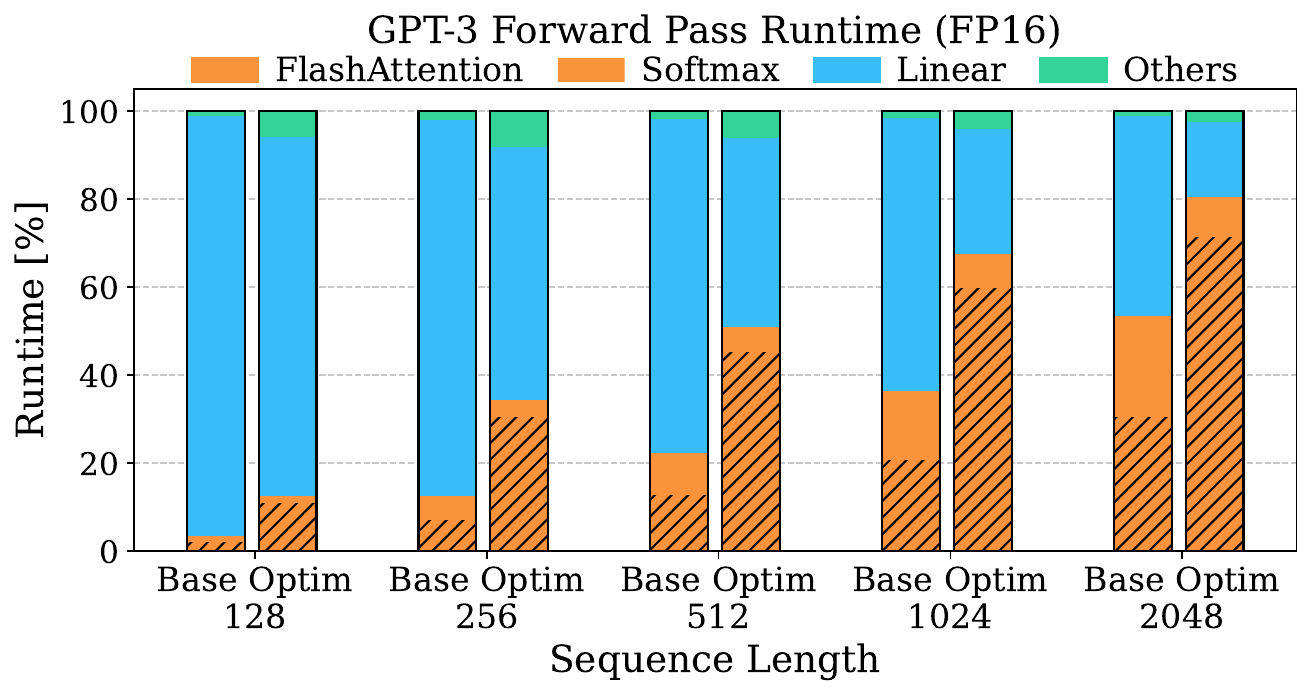}
\caption{Runtime breakdown for GPT-3 on a RISC-V multi-cluster platform~\cite{potocnik_optimizing_2024}. For each sequence length, the left bar shows unoptimized \gls{gemm} results, while the right bar reflects optimized \gls{gemm} results.}
\label{image:introduction}
\end{figure}

Accelerating Softmax poses challenges due to its non-linear, non-pointwise nature and its reliance on a transcendental function, i.e. the exponentiation. The low arithmetic intensity of Softmax constrains parallelism and processing efficiency, a limitation that becomes more pronounced as \gls{gemm} latency decreases with acceleration. For example, the runtime breakdown for BERT on Volta GPU from Steven et al.~\cite{stevens_softermax_2021} shows that Softmax contributes more than 30\% for long sequences. Moreover, as shown in \autoref{image:introduction}, deployment of GPT3-XL on the RISC-V multi-cluster platform~\cite{potocnik_optimizing_2024} reveals that Softmax contributes to 30\% of the runtime prior to \gls{gemm} operator acceleration, and 70\% afterwards for sequence length of 2048.

The use of large accelerators is justified for \glspl{gemm}, which constitute the majority of a Transformer's workload in terms of number of operations. However, allocating considerable silicon area for Softmax acceleration is sub-optimal, as it represents only a small portion of the overall computational workload. Hence, addressing these challenges necessitates innovative solutions that optimize the Softmax function with minimal area costs while preserving accuracy. 

Existing software-level optimizations~\cite{kim_i-bert_2021} often fall short of delivering the accuracy, performance and efficiency improvements necessary for large-scale or low-power deployments. Although hardware accelerators improve performance and energy efficiency, they typically lack flexibility due to their dependence on fixed-function datapath, and often lead to considerable integration area overhead in existing systems. Moreover, achieving accuracy parity often necessitates retraining~\cite{stevens_softermax_2021}~\cite{islamoglu_ita_2023}—a technique that is undesirable, and often impractical for \glspl{llm}. 

In contrast, the growing adoption of the open and extensible RISC-V \gls{isa} to design domain-specialized programmable compute units offers a promising approach for addressing these limitations. In this work, we identify the exponential function as the primary computational bottleneck in Softmax computation, and we accelerate it on RISC-V-based systems, with low hardware overhead, through specialized \gls{isa} extensions. We demonstrate significant performance and energy efficiency gains in accelerating Transformer inference at native precision (Bfloat16), without compromising model accuracy, and without jeopardizing area and power consumption of the compute system. The contributions of this paper are:
\begin{itemize}

\item We design a custom arithmetic block for accelerating the exponential function on \glsentrylong{bf16} data, integrate it  with ultra-low overhead in the \gls{fpu} of programmable RISC-V processors of an octa-core compute cluster, and extend their \gls{isa} with a custom EXP instruction;

\item We implement the cluster using GlobalFoundries \qty[detect-all=true]{12}{\nano\meter} technology down to silicon-ready design and demonstrate that our solution incurs only a 1.0\% area overhead at the cluster level and a negligible power overhead of 1.8\% on workload with peak FPU utilization that do not exploit EXP, while the energy required to execute exponential operation is reduced by two orders of magnitude;

\item By exploiting the proposed \gls{isa} extensions, we optimize the execution of Softmax in software, demonstrating 162.7$\times$ latency reduction and 74.3$\times$ less energy consumption compared to a non-optimized kernel on the baseline cluster and achieving 1.4$\times$ better area efficiency and 7.4$\times$ lower power consumption compared to state-of-the-art Softmax accelerators, as detailed in \autoref{subsec:soa}. Moreover, we integrate our fully optimized Softmax kernel into FlashAttention-2, showing 8.2$\times$ performance and 4.1$\times$ energy efficiency improvements;

\item We scale up the proposed cluster to a 16-cluster system to evaluate the acceleration capabilities of our solution on end-to-end execution of pre-trained, un-tuned Transformers. We benchmark models such as GPT-2, GPT-3, and Vision Transformers (ViT), achieving up to 5.8$\times$ latency reduction and up to 3.6$\times$ less energy consumption compared to a baseline system without the proposed optimizations. Notably, these gains are achieved without re-training and with an accuracy loss of less than 0.1\%.

\end{itemize}

%% file: TEXT/02_Related_Work.tex
\section{Related Work}

Optimization techniques for Softmax can be broadly divided into two categories: workflow scheduling and computational approximations. Workflow scheduling techniques such as FlashAttention\cite{dao_flashattention_2022} and FlashAttention-2 \cite{dao_flashattention-2_2023} enhance data reuse through a tiling technique, thereby improving both memory efficiency and parallelism. While these techniques do not directly target optimizing the Softmax computation kernel, they are complementary to the kind of optimizations explored in this work, which focus specifically on improving the efficiency of the computation kernel itself.

Computational approximations target the core Softmax operations: exponentiation and division. Full-precision exponentiation units based on iterative methods like Taylor series~\cite{van_der_hoeven_fast_2024} and \gls{cordic}~\cite{chen_hgh-cordic_2024} offer accuracy but suffer from slow convergence, hence, long latency and high implementation costs. \gls{lut}-based methods \cite{sun_high_2018} pre-compute values for faster computation but face scalability challenges due to high memory usage. Piecewise linear approximations \cite{dong_plac_2020} balance accuracy and efficiency but require input preprocessing, which can introduce additional overhead. Schraudolph’s method~\cite{schraudolph_fast_1999} achieves fast exponential performance but is limited by accuracy. Division, a key component of Softmax normalization, further adds complexity. Methods like log-sum-exp~\cite{zhu_efficient_2020} eliminate the need for division at the cost of logarithm computations. Alternatively, a single division can compute the reciprocal of the denominator, which is then multiplied by all values for normalization~\cite{stevens_softermax_2021}.

Recently, hardware accelerators for Softmax have emerged, particularly for Transformers. Most accelerators \cite{zhu_efficient_2020, stevens_softermax_2021, koca_hardware-efficient_2023, kim_hardware-efficient_2024, xia_hyft_2024} rely on fixed-point approximations for exponentiation and division, enabling efficient circuitry but complicating precision handling due to conversions from floating-point or integer formats. Methods like \cite{yu_nn-lut_2022, wang_sole_2023} overcome this by operating directly on integer or floating-point formats without fine-tuning, while \cite{stevens_softermax_2021, liu_consmax_2024} are tailored specifically for 8-bit quantized networks but rely on fine-tuning.

To achieve a better balance between flexibility and efficiency, we design a custom arithmetic block for the fast execution of the exponential function on \glsentryshort{bf16} data, a widely adopted precision for Transformers~\cite{burgess_bfloat16_2019}, based on an enhanced version of Schraudolph’s method. We integrate it into RISC-V processors within a parallel compute cluster through lightweight custom \gls{isa} extensions, providing a detailed description and evaluation of the proposed approach, including accuracy and design tradeoffs, while demonstrating significant acceleration speed-ups, minimal implementation costs, and negligible accuracy loss. As highlighted in \autoref{subsec:soa}, our approach offers superior efficiency and flexibility compared to state-of-the-art Softmax accelerators, enabling end-to-end execution of Transformers without the need for fine-tuning.

%% file: TEXT/03_Background.tex
\section{Background}
\subsection{Snitch Cluster}
\label{sec:snitch_cluster}
\begin{figure}[t]
\centering
\includegraphics[width=\linewidth]{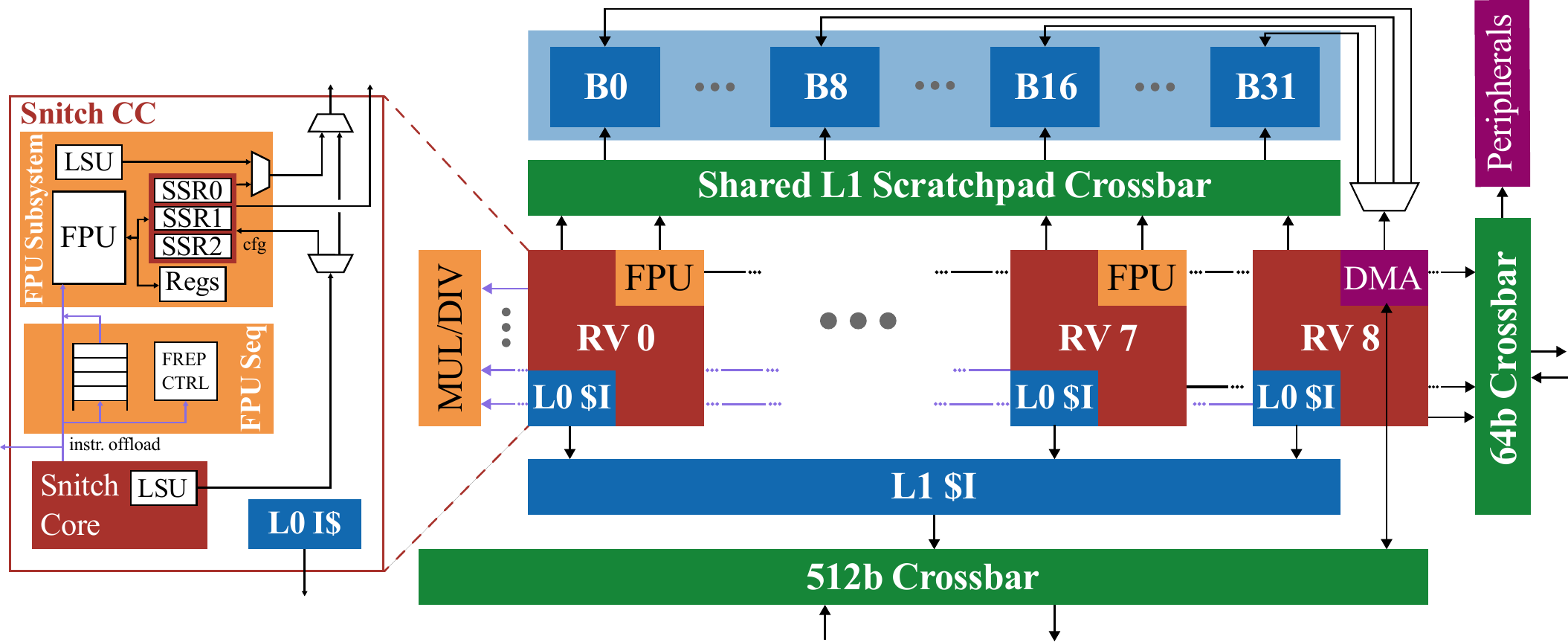}
\caption{Architecture of the RISC-V compute cluster with \gls{isa} extension FREP and SSR~\cite{zaruba_snitch_2021}.}
\label{image:sntich}
\end{figure}
\autoref{image:sntich} illustrates the architecture of the Snitch cluster\cite{zaruba_snitch_2021} that we use as a baseline. The Snitch cluster is an energy-efficient compute architecture designed for high-performance workloads. It integrates eight RISC-V RV32IMAFD cores, each paired with a tightly coupled 64-bit \gls{simd}-capable \gls{fpu} supporting a wide range of data formats (FP64 to FP8, including BF16) and a private L0 instruction cache.

A 128 KiB, 32-banked \gls{spm} is shared across the cluster, connected via a single-cycle logarithmic interconnect that delivers high-bandwidth, low-latency data access. A dedicated \gls{dma} control core facilitates asynchronous data transfers between the \gls{spm} and external memory systems (e.g., HBM2E or other clusters), achieving bandwidths of up to \qty[detect-all=true]{512} {\bit\per{cycle}}. To ensure efficient data movement, the hierarchical interconnect incorporates a 512-bit wide crossbar for L1 instruction cache and data access and a 64-bit crossbar for peripheral communication.

The architecture also supports advanced \gls{isa} extensions, including \texttt{FREP} (Floating-Point Repetition)\cite{zaruba_snitch_2021} for hardware loops and \texttt{SSR} (Stream Semantic Register) \cite{schuiki_stream_2021} for managing data access with minimal software overhead. The \texttt{FREP} instruction configures the FPU sequencer to automatically repeat and autonomously issue the next \(n\) floating-point instructions to the FPU. The \texttt{SSR} extension allows the configuration of up to three memory streams with affine address patterns, effectively eliminating explicit memory operations.

\subsection{FlashAttention}



A Transformer consists of multiple blocks, each containing a multi-head attention (MHA) module and a feed-forward module. In MHA, token vectors are projected through query (Q), key (K), and value (V) matrices, with attention computed as $\text{Softmax}((QK^\top)/\sqrt{d_k})V$. FlashAttention\cite{dao_flashattention_2022} optimizes this computation by dividing Q, K, and V into blocks that can be efficiently processed in fast SRAM memory, thereby reducing costly HBM accesses. FlashAttention-2\cite{dao_flashattention-2_2023} further enhances performance through optimized memory layouts and aggressive operator fusion.

For numerical stability, we adopt the Softmax function with maximum subtraction:
$$
\text{Softmax}(x_i) = \frac{\exp(x_i - \max(\mathbf{x}))}{\sum_j \exp(x_j - \max(\mathbf{x}))}
$$
, which requires storing the complete attention matrix and performing row-wise operations. To address this limitation, FlashAttention introduces \textit{partial} Softmax, which processes blocks incrementally while maintaining running statistics (maximum values and exponential sums). For each new block, these statistics are updated and used to compute partial results, enabling numerically equivalence  to standard Softmax while significantly reducing memory overhead. This online computation approach not only ensures numerical stability but also eliminates the need to materialize the full attention matrix in memory.

\subsection{Execution Model on Snitch Cluster}\label{subsection:exection_softmax}

\textbf{Baseline Softmax:} The baseline kernel is written in C without leveraging Snitch's extended \gls{isa} (FREP, SSR and SIMD). Data is transferred via \gls{dma} from HBM to the local \gls{spm} with double buffering to mask data marshalling latency while the eight Snitch cores process sequences in parallel.  The division in Softmax is performed by the FPU's division block, and the exponential function is based on \texttt{math.h} library and uses a piecewise polynomial approximation method with software \glspl{lut}.

\textbf{Baseline FlashAttention-2:} 
Following the approach in~\cite{potocnik_optimizing_2024}, we adapt FlashAttention-2 to the Snitch cluster architecture with an optimized tiling strategy. The implementation first loads a Q tile to \gls{spm} via \gls{dma}, then iteratively transfers and processes corresponding K and V tiles. To maximize throughput, we employ double buffering for efficient overlap between memory transfers and computation. The tile size is optimized based on \gls{spm} capacity under double buffering constraints. Within each tile, both \gls{gemm} and partial Softmax computations are parallelized across the cluster cores. The partial Softmax computation is parallelized by having the eight cluster cores simultaneously compute multiple row statistics. The \gls{gemm} implementation leverages Snitch's specialized instruction-level optimizations as detailed in \cite{potocnik_optimizing_2024}, which serves as the foundation for all \gls{gemm} operations in this work.

\subsection{Exponential Approximation Algorithm}\label{subsec:exp_approximation}
For efficient exponential computation, we adopt Schraudolph's method \cite{schraudolph_fast_1999}, which exploits the memory arrangement of floating-point numbers to approximate \(e^x\) with few basic operations. The input \(x\) is scaled to the base-2 domain as \(x' = x / \ln(2)\), then decomposed into integer and fractional parts: \(\mathrm{int}(x') = \lfloor x' \rfloor\) and \(\mathrm{frac}(x') = x' - \lfloor x' \rfloor\). The approximation is reconstructed as \(\exp(x) \approx 2^{\mathrm{int}(x')} \cdot \left(1+\mathrm{frac}(x')\right)\). Based on the method proposed by Belano et al.~\cite{belano_flexible_2024}, to enhance accuracy, the fractional term \(\left(1+\mathrm{frac}(x')\right)\) is replaced with a polynomial \(P(\mathrm{frac}(x'))\), yielding:

\vspace{-0.2em}
\begin{equation}\label{exp_correction}
    \exp(x) \approx 2^{\mathrm{int}(x')} \cdot \left(1 + P\left(\mathrm{frac}(x')\right)\right).
\end{equation}

To better approximate \(2^{\mathrm{frac}(x)}\), the interval \([0,1)\) is split into two equal-length partitions, determined by the most significant bit of the mantissa. For each partition, a polynomial in the form \(ax(x+b)\) is applied:
\vspace{-0.2em}
\begin{equation}
P(x) =
\begin{cases}
\alpha x \left(x + \gamma_1 \right), & x \in [0, 0.5), \\
\mathrm{not}\left(\beta \, \mathrm{not}(x) \cdot \left(x + \gamma_2 \right)\right), & x \in [0.5, 1).
\end{cases}
\end{equation}

Here, \(\alpha\), \(\beta\), \(\gamma_1\), and \(\gamma_2\) are optimized for minimal error, with \(1 - x\) approximated by \(\mathrm{not}(x)\) for hardware efficiency. Adjustments to \(\gamma_1\) and \(\gamma_2\) account for fixed-point arithmetic constraints. Parameters \(\alpha = 0.21875\), \(\beta = 0.4375\), \(\gamma_1 = 3.296875\) and \(\gamma_2 = 2.171875\) are derived via a heuristic Monte Carlo optimization with $10^6$ trials by Belano et al.~\cite{belano_flexible_2024}  to minimize the error between the true exponential function and its approximation.

%% file: TEXT/04_HW_SW_Optimizations_v2.tex
\section{Methods}

\subsection{EXP Custom Arithmetic Block}
\label{sec:exp_block}
\begin{figure}[t]
\centering
\includegraphics[width=\linewidth]{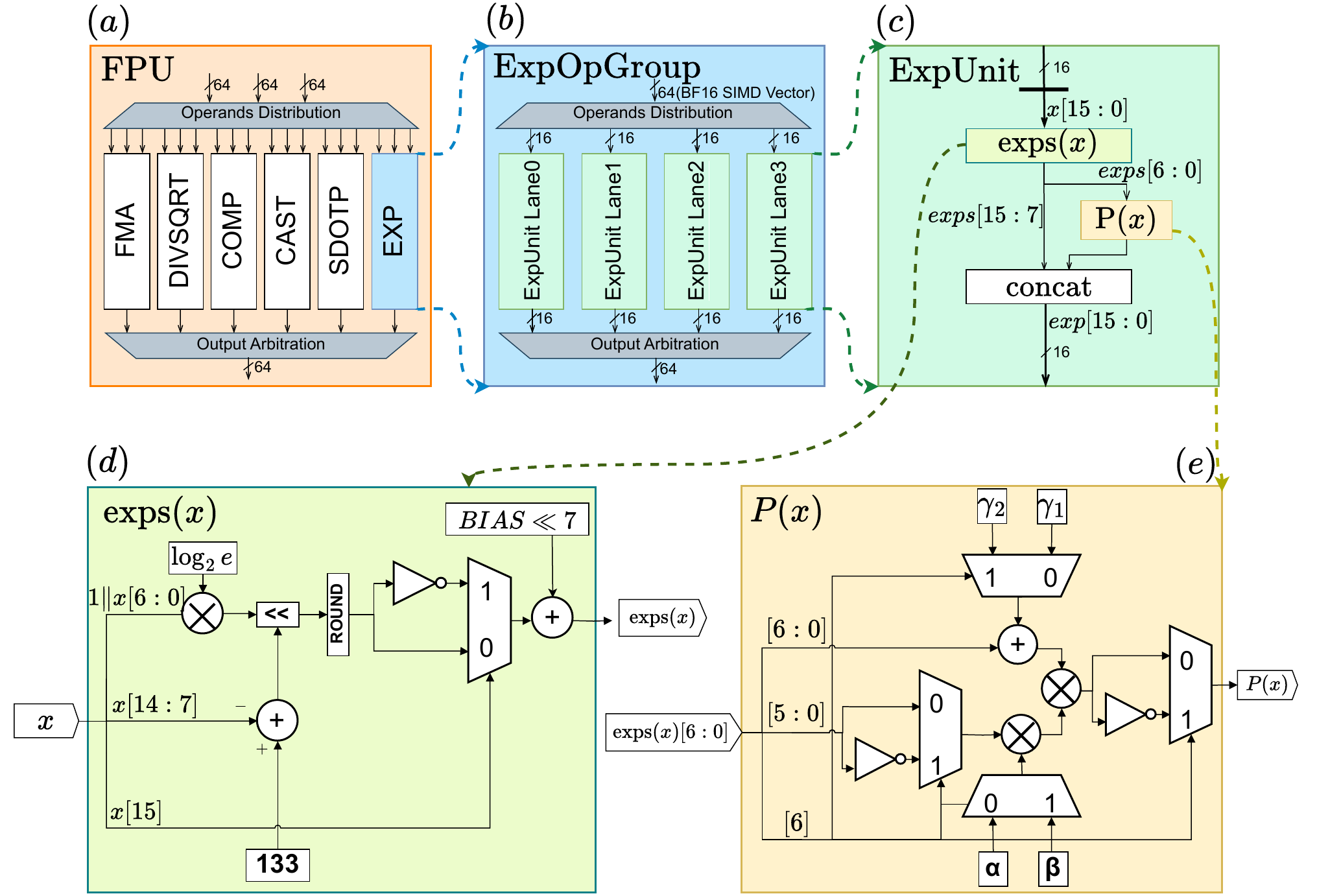}
\caption{Block diagram of (a) the extended \gls{fpu}, (b) the \textit{ExpOpGroup}, (c) the \textit{ExpUnit}, (d) the $exps(x)$ stage, and (e) the $P(x)$ stage.} 
\label{datapath}
\end{figure}


\autoref{datapath}c shows the proposed arithmetic block to efficiently compute the approximation of exponential function on Bfloat16 data. This block is structured around the algorithm introduced in \autoref{subsec:exp_approximation} and consists of two cascaded stages as described above: \textbf{$exps(x)$}, which implements the Schraudolph's method in hardware, and the subsequent \textbf{$P(x)$}, which performs the mantissa correction for improved precision. 

At the input of the \textbf{$exps(x)$} stage (shown in \autoref{datapath}d), the data in Bfloat16 format is decomposed into its sign, exponent, and mantissa bits, with the implicit leading 1 appended to the latter. Next, the mantissa is multiplied by the precomputed constant ($\log_2 e$), and the result of this multiplication is then shifted 
by an amount equal to the difference between the exponent of the argument and the maximum exponent after which the exponential function is guaranteed to overflow (133 in the case of BFloat16 numbers).
Then, the first 15 bits of the shifted mantissa are selected and appropriately rounded to maintain precision, and finally, if no overflow occurs, the result is obtained by appending a leading zero (the sign bit of the result) to the rounded mantissa and adding the bias to the new exponent. If an overflow or infinity is detected, the output is assigned to either \( \infty \) or \( 0 \), depending on whether the argument is positive or negative. For subnormal values, the data is flushed to zero following BFloat16 simplifications relative to IEEE-754 behaviour \cite{burgess_bfloat16_2019}.

\vspace{-0.1em}
The second stage of the exponential computation, \textbf{$P(x)$}, corrects the mantissa component of the approximation. First, the MSB of the mantissa determines whether the input falls within \([0, 0.5)\) or \([0.5, 1)\), selecting the appropriate polynomial branch. In the first branch, corresponding to \(x \in [0, 0.5)\), the polynomial \(\alpha x (x + \gamma_1)\) is evaluated directly using fixed-point arithmetic. For \(x \in [0.5, 1)\), the computation proceeds with \(\mathrm{not}(\beta\, \mathrm{not}(x) \cdot (x + \gamma_2))\), where the bitwise complement operation approximates the evaluation of \(1 - x\). 

Finally, the output of the EXP block is obtained by concatenating the corrected mantissa from the \textbf{$P(x)$} stage with the sign and exponent fields from the \textbf{$exps(x)$} stage.

\subsection{Snitch \gls{isa} Extension and Microarchitecture}




To exploit the fast exponentiation of Bfloat16 data enabled by the EXP block described in \autoref{sec:exp_block} while preserving software programmability, we integrate the arithmetic block into an open-source, modular, energy-efficient multiformat \gls{fpu}~\cite{bertaccini_minifloat-nn_2022} for RISC-V processors.  The target \gls{fpu} already supports a wide range of floating-point operations, which are organized into specific multi-format modules that can be enabled at design-time through parameters. For our evaluation, it integrates \texttt{FMA} (fused multiply-add), \texttt{DIVSQRT} (division and square root), \texttt{COMP} (comparison), \texttt{CAST} (conversion), and \texttt{SDOTP} (dot product) modules. We extend it with a new dedicated single-format module, namely \textit{ExpOpGroup}. 

The new operation group takes as input a single $N$-bit SIMD vector containing Bfloat16 elements and produces a single $N$-bit SIMD output vector, as illustrated in~\autoref{datapath}b. Depending on the data-width of the \gls{fpu}, which is configurable at design time, the \textit{ExpOpGroup} integrates $k=N$-bit$/16$-bit \textit{ExpUnit} lanes. This is preceded by additional logic that segments the input SIMD vector into $k$ 16-bit elements and distributes them to the \textit{ExpUnit}s. To meet the timing requirements of the processor that integrates the \gls{fpu} enhanced with the proposed block, the \textit{ExpUnit} includes a configurable number of pipeline stages, which can be utilized for retiming.

Furthermore, the extended \gls{fpu} is integrated into the micro-architecture of the Snitch cores of the parallel compute cluster introduced in \autoref{sec:snitch_cluster}. Since the Snitch core supports double-precision instructions, the \gls{fp} register file contains 32 64-bit wide registers and its datapath is 64-bit wide. At the interface, the \gls{fpu} accepts three 64-bit input operands and produces one 64-bit output per cycle. This configuration allows the packing of four Bfloat16 operands into a single SIMD 64-bit register. As illustrated in~\autoref{datapath}c, the \textit{ExpOpGroup} is parameterized with four \textit{ExpUnit} lanes, each equipped with one level of pipeline registers to streamline processing. This design allows for the completion of a single exponentiation operation in two cycles, while still permitting back-to-back operations without stalling, thereby maintaining a peak throughput of four Bfloat16 exponentiation operations per cycle.

To enable this operation in software, we extend the Snitch's RISC-V \gls{isa} with two domain-specific instructions, namely \texttt{FEXP} and \texttt{VFEXP}. The first instruction is designed for scalar Bfloat16 operations, activating only one \textit{ExpUnit} in the micro-architecture, while \texttt{VFEXP} performs a packed-SIMD exponential computation that fully utilizes the underlying micro-architecture's capabilities. Both instructions execute with a latency of two clock cycles in the Snitch core. As shown in \autoref{tab:riscv_encodings}, the instruction formats use \texttt{rd} (destination register) and \texttt{rs1} (source register) as 5-bit fields to address the \gls{fpu}’s 32$\times$64 register file, with the most significant bit of the entire instruction distinguishing between scalar and packed-\gls{simd} operations. The Snitch core’s decoder and \gls{fpu} subsystem are updated to support these instructions and seamlessly activate the \textit{ExpOpGroup} in the \gls{fpu}.

\subsection{Optimized Softmax Kernel}
\begin{table}[t]
\centering
\caption{Snitch RISC-V Encodings for FEXP and VFEXP}
\label{tab:riscv_encodings}
\setlength{\tabcolsep}{4pt}
\begin{tabular}{lc}
\toprule
\textbf{Format}        & \textbf{Encoding (32-bit)}        \\ 
\midrule
\texttt{\textbf{FEXP} \ rd, rs1}  & \texttt{001111100000\{rs1\}000\{rd\}1010011} \\ 
\texttt{\textbf{VFEXP} rd, rs1} & \texttt{101111100000\{rs1\}000\{rd\}1010011} \\ 
\bottomrule
\end{tabular}
\end{table}
\begin{figure}[tb]
\centering
\includegraphics[width=\linewidth]{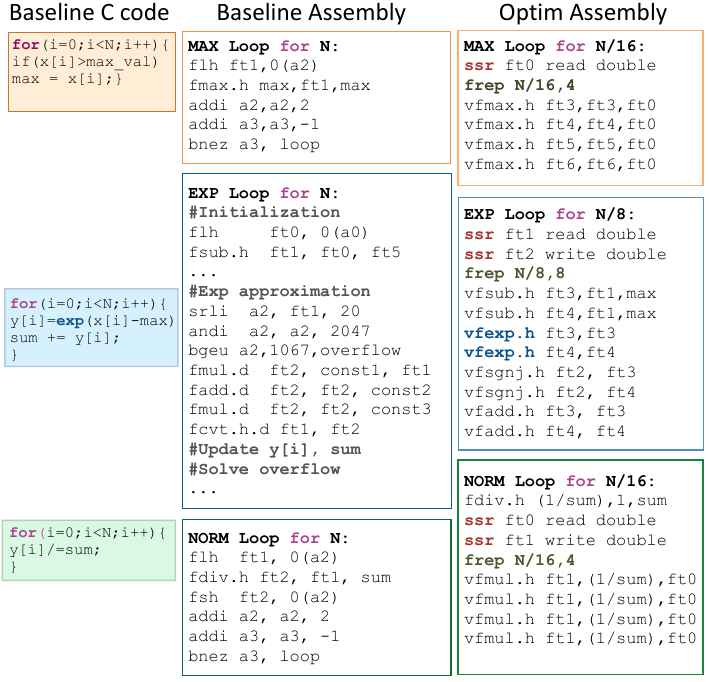}
\caption{Code comparison of Baseline and Optimized Softmax implementations. 
Baseline Softmax uses a piecewise polynomial approximation with software LUTs for the exponential (EXP) function, explicitly handling overflow to infinity and subnormals. The notation \texttt{frep n\_frep, n\_instr} represents a loop executing the following \texttt{n\_instr} instructions for \texttt{n\_frep} iterations. All \texttt{v} instructions in the code are packed-SIMD operations.}
\label{image:code}
\end{figure}






To speed up the execution of the Softmax function on the enhanced Snitch cluster, we develop optimized software routines that exploit the underlying ISA of the Snitch cores, including the designed \texttt{VFEXP} instruction. 
%
Since the maximum value is required for the exponentiation step and must be computed by looping over each row of the resulting \( QK^T \) matrix, we construct the loop using the \texttt{FREP} instruction. As shown in \autoref{image:code}, targeting BF16 data and leveraging the core's 64-bit datapath, we utilize the \texttt{VFMAX} instruction in MAX step to balance computation and load costs - processing 4 \gls{simd} operations per 64-bit data load. To streamline data loads and keep the datapath fully utilized, we exploit an \texttt{SSR}. 

The results are then forwarded to the exponentiation step (EXP), where we maintain a similar kernel structure using \texttt{FREP} and \texttt{SSR} for efficient data loading. In this phase, we leverage the \texttt{VFEXP} instruction, which performs the exponentiation of a \gls{simd} vector with 4 elements in 2 cycles. In contrast, the baseline kernel, described in \autoref{subsection:exection_softmax}, computes the exponentiation in software with a latency of 319 cycles per BF16 item. For each computed exponential, we also accumulate the sum using \texttt{VFADD} within the same \texttt{FREP-SSR} loop. 

Finally, we optimize the normalization step (NORM) by calculating $1/\text{sum}$ outside the loop and performing a point-wise scaling operation with a \texttt{VFMUL} instruction. Overall, these optimizations achieve 1.5 instructions/output, 2.125 cycles/output while also benefiting from loop unrolling advantages, significantly outperforming the baseline implementation, which requires 56 instructions/output, 360 cycles/output.

\subsection{Optimized FlashAttention-2 Kernel}We optimize the partial Softmax part of the FlashAttention-2 kernel, which follows steps analogous to standard Softmax but performs them over multiple tiles. The optimization methods including \texttt{FREP}, \texttt{SSR} and \gls{simd} instructions (\texttt{VFEXP}, \texttt{VFMAX}, \texttt{VFSUB}, \texttt{VFMUL}) are employed in the same manner for the partial Softmax in FlashAttention-2 for partial MAX, EXP, and NORM.

%% file: TEXT/05_Evaluation.tex
\section{Evaluation and Results}
\begin{table}[t]
\centering
\caption{Accuracy for GPT-2 and ViT Models}
\label{table:accuracy}
\begin{tabular}{c c c c c c}
\toprule
\textbf{Model} & \textbf{Dataset} & \textbf{Metric} & \textbf{FP32} & \textbf{BF16} & \textbf{BF16 EXP}  \\ 
\midrule
\multirow{2}{*}{GPT-2} & WikiText & Perplexity ($\downarrow$) & 37.4  & 37.8 &  37.8 \\ 
                       & ArcEasy  & Accuracy  ($\uparrow$) & 43.8   & 42.9  & 43.7 \\ 
\midrule
\multirow{2}{*}{ViT-B}   & ImageNet & Accuracy ($\uparrow$) & 80.3 & 80.3 & 80.3 \\ 
                       & CIFAR-10 & Accuracy  ($\uparrow$)   & 98.5 & 98.5 & 98.5 \\ 
\bottomrule
\end{tabular}
\end{table}

\subsection{Accuracy Analysis}
\label{subsec:accuracy}
Following Belano et al. \cite{belano_flexible_2024}, the proposed exponentiation algorithm achieves a mean relative error of 0.14\% and a maximum relative error of 0.78\% with respect to glibc’s implementation. Building upon this, we evaluate the accuracy of our exponential implementation using 
 pre-trained GPT-2 Small and ViT Base models. For GPT-2 Small, perplexity is measured on WikiText-2, and accuracy is measured on ARC Easy. For ViT Base, accuracy is evaluated on ImageNet-1K and CIFAR-10. Comparisons are made against FP32 precision, native BF16 casting, and BF16 casting with our optimized EXP implementation, which employs a software-simulated Schraudolph algorithm.

As shown in \autoref{table:accuracy}, BF16 casting has minimal impact on the accuracy of the models. Moreover, our proposed EXP replacement demonstrates negligible differences compared to standard BF16 casting. These findings validate the proposed EXP approach as an efficient and accurate method for exponential computation, preserving model accuracy. Notably, this analysis demonstrates that Transformer models can be directly cast to BF16 without the need for re-training or fine-tuning, further highlighting the practicality of our approach.


\subsection{Physical Implementation}
\begin{figure}[tb]
\centering
\includegraphics[width=0.95\linewidth]{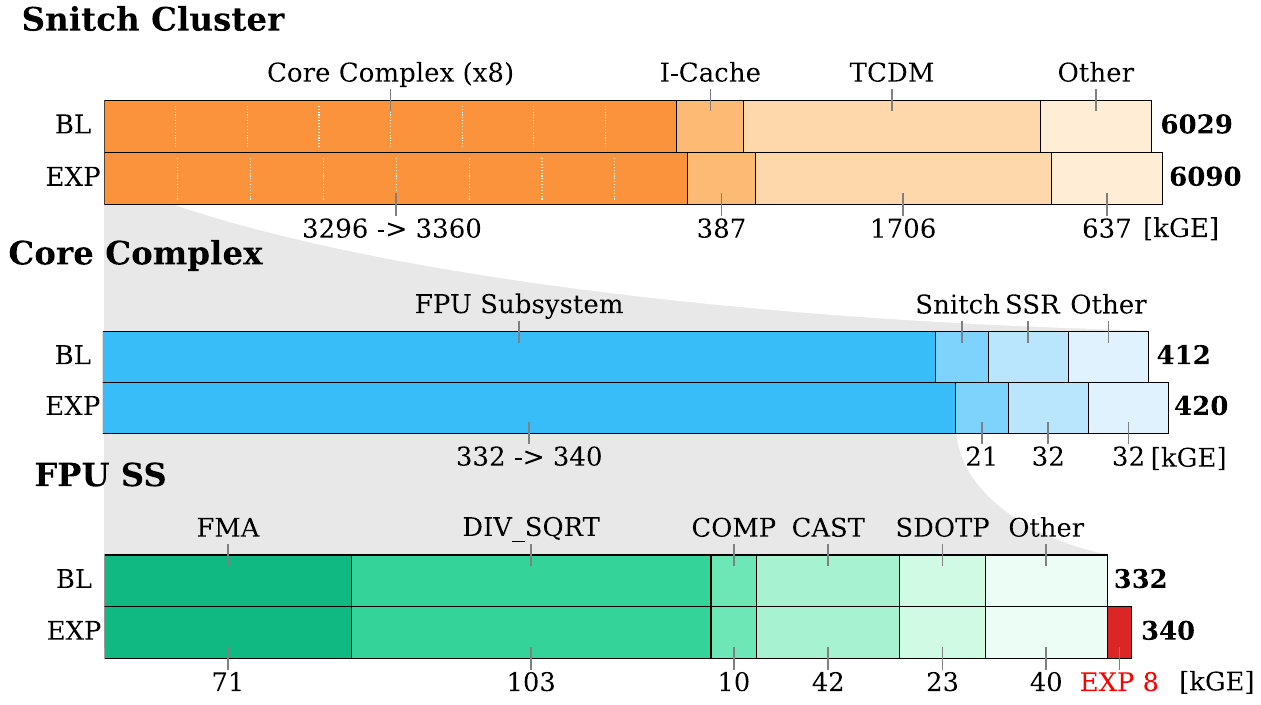}
\caption{Area breakdown of the Snitch cluster. BL: Baseline, EXP: Extended FPU with the EXP block.} 
\label{area_breakdown}
\end{figure}

\begin{table}[t]
\centering
\caption{Energy Per Operation for \gls{gemm} and EXP}
\label{tab:power_comparison}
\begin{tabular}{ccc}
\toprule 
\textbf{Energy/Op [pJ/Op]} & \textbf{Snitch Baseline} & \textbf{\gls{isa} Extended} \\ 
\midrule
\textbf{\gls{gemm}} & 3.96 & 4.04 \\ 
\textbf{EXP} & 3433 & 6.39 \\ 
\bottomrule
\end{tabular}
\end{table}

\begin{figure*}[t]
    \centering
    \includegraphics[width=\textwidth]{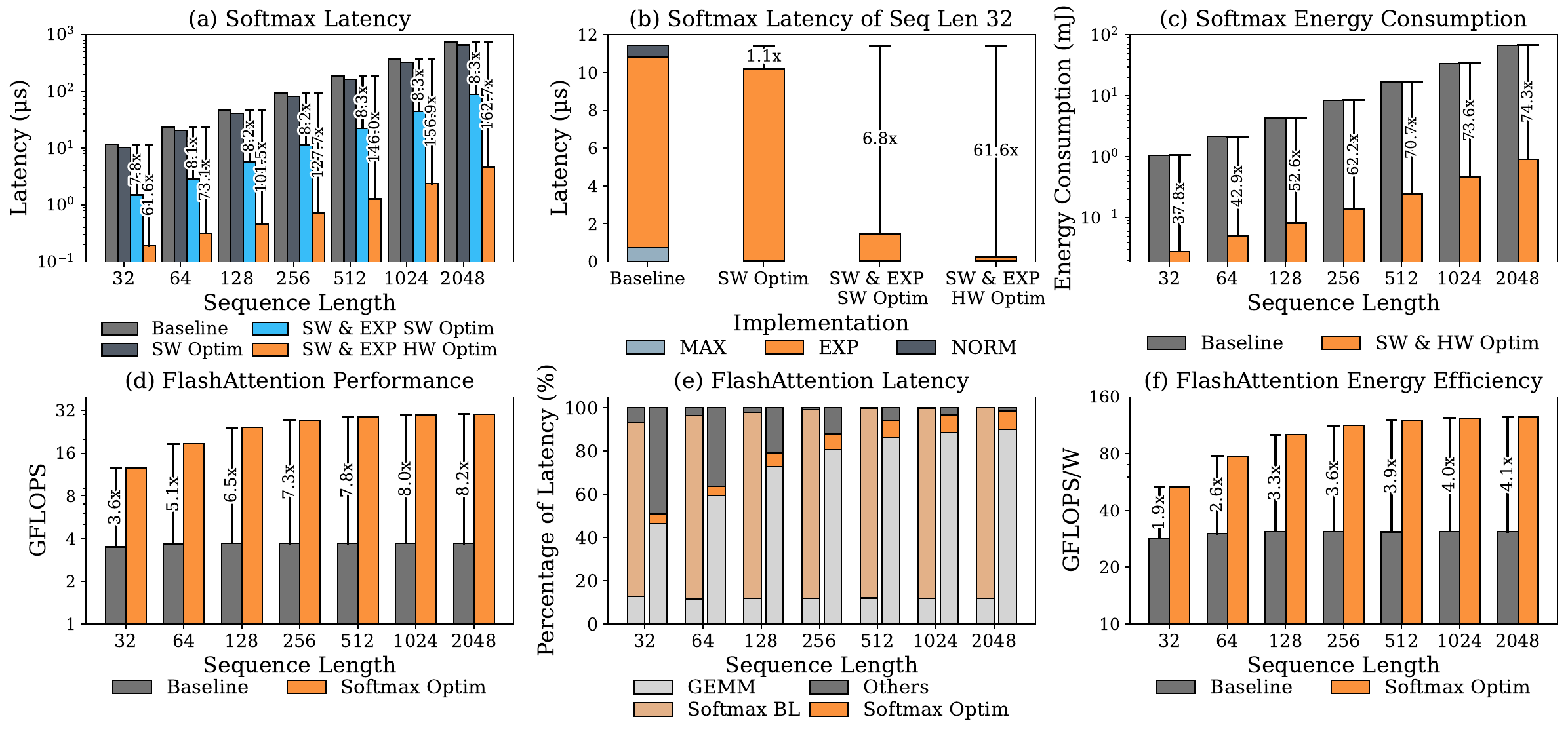}
    \caption{
Performance, latency, and energy analysis for Softmax and FlashAttention-2 kernels. }
\label{fig:double_column_example}
\end{figure*}

We performed synthesis and place\&route for one Snitch cluster with eight cores, 128 KiB of TCDM, and 8 KiB of instruction cache. Synthesis and implementation results are gathered with Synopsys' Fusion Compiler 2022.03 for GlobalFoundries 12nm FinFET technology.

For timing analysis, we constrained the design to \qty[detect-all=true]{1}{\giga\hertz}. With the addition of the exponentiation block, the Snitch cluster achieved \qty[detect-all=true]{1.15}{\giga\hertz} under typical conditions (TT/\qty[detect-all=true]{0.8}{\volt}/\qty[detect-all=true]{25}{\celsius}) without introducing any new critical paths. Under worst-case conditions (SS/\qty[detect-all=true]{0.72}{\volt}/\qty[detect-all=true]{125}{\celsius}), the design reached up to \qty[detect-all=true]{941}{\mega\hertz}.


For the area analysis, we evaluated the cluster, core complex (comprising the integer core and the FPU subsystem), and FPU subsystem (SS), as shown in \autoref{area_breakdown}. At the cluster level, the total area increased by 1.0\% compared to the baseline Snitch cluster due to the increase in the area of the eight core complexes. At the core complex level, the FPU SS exhibited a 1.9\% area increase relative to the baseline. Within the FPU SS, the addition of the EXP block accounted for 8 kilo Gate Equivalents\footnote{ One Gate Equivalent (GE) represents the area of a minimum-sized two-input NAND gate, which is \qty[detect-all=true]{0.121}{\micro\meter^2} in GF 12nm technology.} (kGE), corresponding to a 2.3\% increase in the area of SS.

To measure power, we performed parasitics-annotated gate-level netlist simulations using Synopsys’ PrimeTime 2022.03 under typical conditions (TT/\qty[detect-all=true]{0.8}{\volt}/\qty[detect-all=true]{25}{\celsius}). In \autoref{tab:power_comparison}, \gls{gemm} kernels (48×48, 85\% FPU utilization) are compared between the baseline Snitch and \gls{isa}-extended Snitch. Adding the EXP block increased Snitch cluster’s average power by 1.8\%, with energy per operation rising from 3.96 to \qty[detect-all=true]{4.04}{\pico\joule\per{Op}}.

For EXP implementation, the new EXP instruction was benchmarked against the baseline method (piecewise \gls{lut} with polynomial approximation), which requires 319 cycles per call and has low FPU utilization (6.5\%). The \gls{isa}-extended Snitch core performs exponential calculations in hardware in two clock cycles. During the execution of the EXP kernel, the \gls{isa}-extended Snitch’s average power increased by 2.4$\times$. However, the execution time dropped from \qty[detect-all=true]{319}{{cycles}\per{output}} to \qty[detect-all=true]{0.5}{{cycles}\per{output}}  (with 4-way \gls{simd} instruction \texttt{VFEXP}), reducing the energy from 3433 to \qty[detect-all=true]{6.39}{\pico\joule\per{Op}}.





\subsection{Benchmarks}
\input{TABLEs/Soa_Table}
\textbf{Softmax}: We evaluated four Softmax implementation configurations: the baseline described in \autoref{subsection:exection_softmax}, an optimized version using Snitch’s existing \gls{isa} extensions (\texttt{SW Optim}), a further optimized version incorporating the software-implemented Schraudolph exponential function (\texttt{SW \& EXP SW Optim}), and a final version combining Snitch’s \gls{isa} extensions with hardware acceleration via the EXP instruction (\texttt{SW \& EXP HW Optim}). Performance benchmarks were conducted using ModelSim-2022.3, with the system running at \qty{1}{\giga\hertz}.


In \autoref{fig:double_column_example}a, our final implementation (\texttt{SW \& EXP HW Optim}) achieve up to 162.7$\times$ speedup over the baseline, while software-only optimizations show minimal gains due to the exponential operation bottleneck. The Schraudolph method in software offers some acceleration but is far outperformed by hardware by a factor of 19.6$\times$. \autoref{fig:double_column_example}b demonstrates the negligible impact of MAX and NORM on total latency, with software achieving only a 1.1$\times$ speedup, compared to 61.6$\times$ for combined hardware and software optimizations. Finally, \autoref{fig:double_column_example}c shows energy reductions of up to 74.3$\times$.

\textbf{FlashAttention-2}: We evaluated the FlashAttention-2 kernel on one Snitch cluster with a head dimension of 64 (GPT2 configuration). The results, shown in the second row of \autoref{fig:double_column_example}, highlight several improvements. In \autoref{fig:double_column_example}d, our implementation achieves up to 8.2$\times$ increase in throughput over the baseline. \autoref{fig:double_column_example}e illustrates that Softmax dominates the latency in the baseline, while its contribution is reduced to 6\% in the optimized version. Moreover, the energy efficiency of FlashAttention-2 improves up to 4.1$\times$ with the optimized Softmax as shown in \autoref{fig:double_column_example}f.

\subsection{Scalability Analysis}

\begin{figure}[t]
\centering
\includegraphics[width=0.9\linewidth]{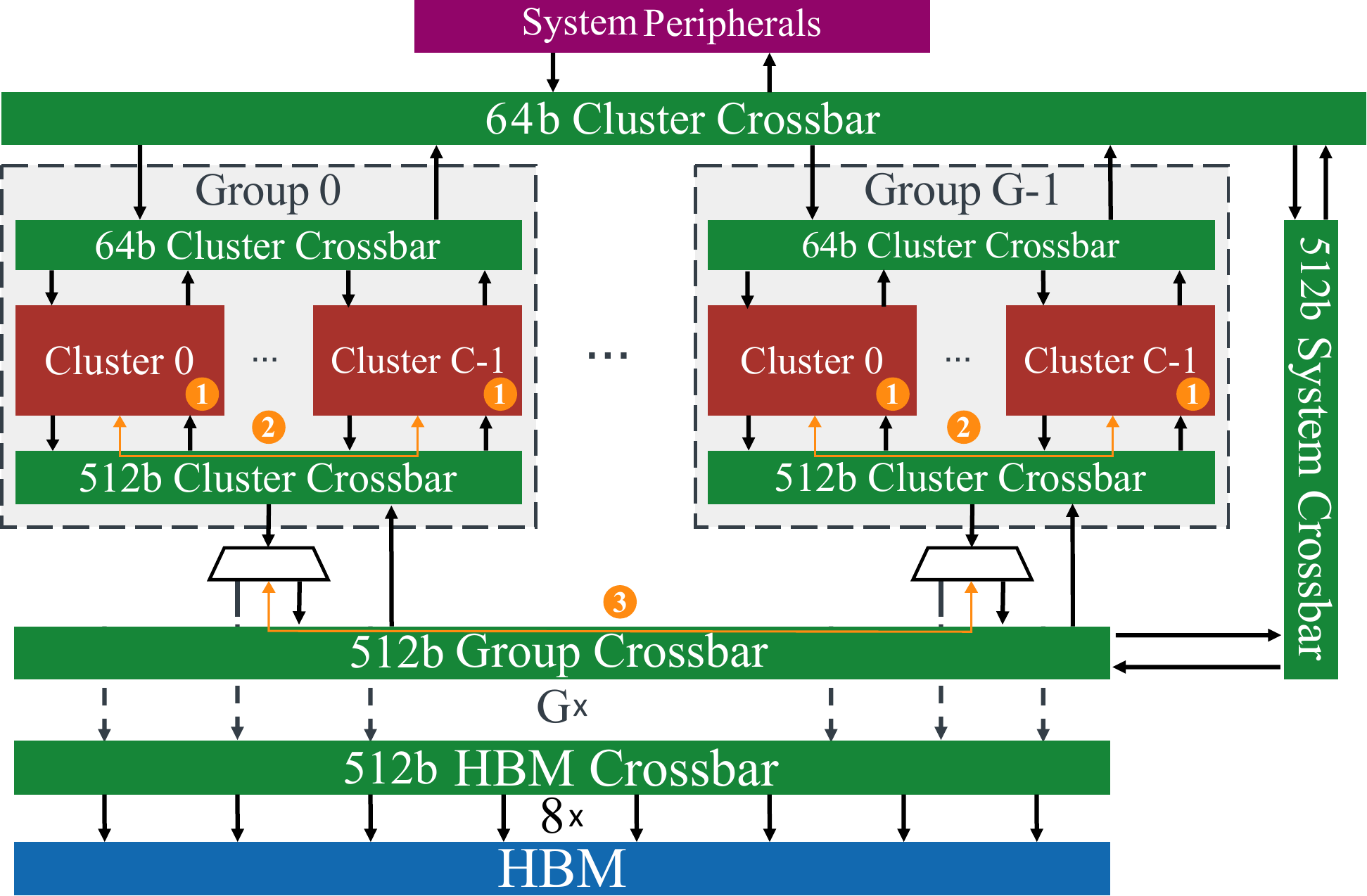}
\caption{Hierarchical multi-cluster architecture with heterogeneous memory interconnect: (1) Cluster-to-SPM interconnect, (2) Inter-Cluster communication, and (3) Inter-Group communication. }
\label{snitch_group}
\end{figure}

The Snitch cluster is designed to scale into a multi-cluster architecture,  silicon-proven in Occamy~\cite{paulin_occamy_2024}. As shown in \autoref{snitch_group}, a group of \(C\) compute clusters is connected by a 64-bit crossbar for fast synchronization and a 512-bit AXI crossbar for high-bandwidth inter-cluster access. Further scaling is achieved by linking \(G\) groups through a group-level AXI crossbar, enabling inter-group communication. Each group also interfaces with eight HBM channels through a wide crossbar, ensuring high-bandwidth access to main memory. 

We benchmark runtime and energy metrics against \cite{potocnik_optimizing_2024} on GPT-2 Small, GPT-3 XL, ViT-Base, and ViT-Huge models. All models are evaluated non-autoregressively on a 16-cluster version of the Occamy system \cite{paulin_occamy_2024}, with sequence lengths of 2048 for GPT models and 197 for ViT models. Following \cite{potocnik_optimizing_2024}, we map each attention head to a single Snitch cluster, loading each Q tile from HBM to SPM via DMA and iteratively transferring and processing the corresponding K and V tiles.

\begin{figure}[t]
\centering
\includegraphics[width=\linewidth]{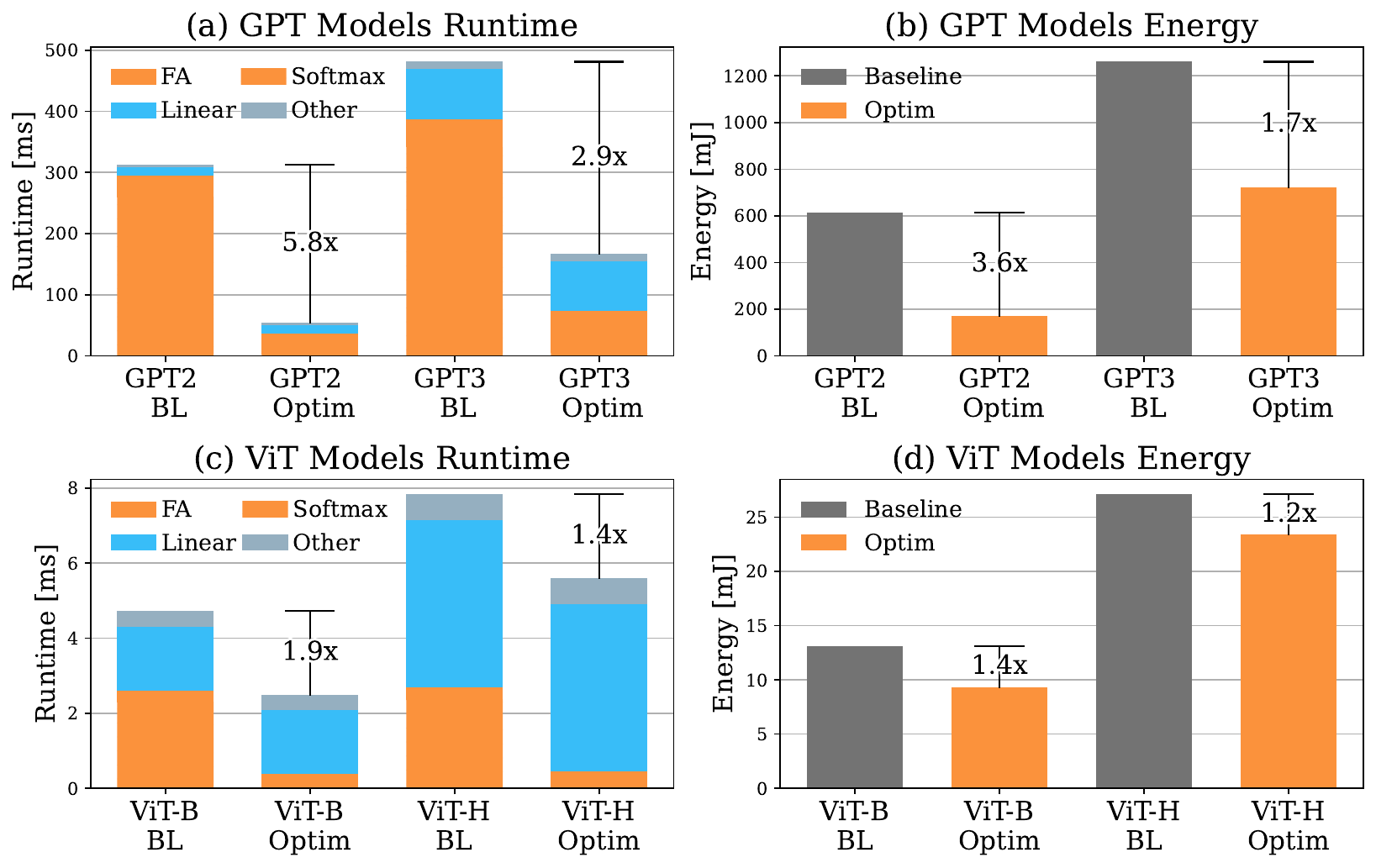}
\caption{Runtime and energy comparison of Softmax-optimized (Optim) system with the baseline (BL) for GPT and ViT models.}
\label{end2end_results}
\end{figure}
As shown in \autoref{end2end_results}, the FlashAttention-2 kernel dominates runtime in the baseline implementation for both GPT and ViT models.  With Softmax optimizations applied, overall runtime improves significantly, achieving speedups of 5.8$\times$, 2.9$\times$, 1.9$\times$, and 1.4$\times$ for GPT-2, GPT-3, ViT-Base, and ViT-Huge, respectively. Similarly, energy consumption decreases substantially, with reductions of 3.6$\times$, 1.7$\times$, 1.4$\times$, and 1.2$\times$ for these models, respectively.

%% file: TABLEs/Soa_Table.tex
\begin{table*}[t]
\centering
\caption{Comparison of State-of-the-Art Softmax Accelerators}
\scriptsize
\resizebox{\textwidth}{!}{%
\begin{threeparttable} 
\begin{tabular}{c c r c c r r r r c}
\toprule
\multirow{2}{*}{\textbf{Ref}} & \multirow{2}{*}{\textbf{Precision}} & \textbf{Accuracy} & \textbf{Evaluated} & \textbf{Tech} & \textbf{Frequency\tnote{*}} & \textbf{Area\tnote{*}} & \textbf{Power} & \textbf{Throughput} & \multirow{2}{*}{\textbf{Strategy}} \\ 

& & \textbf{[MSE]} & \textbf{Model} & \textbf{[\qty[detect-all=true]{}{\nano\meter}]} & \textbf{[\qty[detect-all=true]{}{\giga\hertz}]} & \textbf{[\qty[detect-all=true]{}{\micro\meter^2}]} & \textbf{[\qty[detect-all=true]{}{\milli\watt}]} & \textbf{[\qty[detect-all=true]{}{\giga OPS}]} &  \\

\midrule


Zhu et al.\cite{zhu_efficient_2020} & FX16 & 1.06e-10/2.28e-12\tnote{1} & Transformer-XL & 28 & 2.78/1.64\tnote{1} & 10081/18392\tnote{1} & - & 22.24/13.12\tnote{†,1} & FX16 quant. \\

Koca et al.\cite{koca_hardware-efficient_2023} & FX16 & - & BERT & \textit{FPGA} & - & - & - & - & No fine-tuning \\

Kim et al.\cite{kim_hardware-efficient_2024} & FX8/FX16 & 71.2e-12/4.77e-12\tnote{2} & - & 28 & 3.12/2.5\tnote{2} & 7100/24900\tnote{2} & 22.82/52.46\tnote{2} & 24.96/20\tnote{†,2} &  - \\

Xia et al.\cite{xia_hyft_2024} & FP16/FP32-FX\tnote{3} & - & BERT & \textit{FPGA} & - & - & - & - & Fine-tuning \\

Yu et al.\cite{yu_nn-lut_2022}& INT32/FP16/FP32 & - & RoBERTa, MobileBERT & 7 & 1.5/0.74/0.62\tnote{4} & 1009/498/1134\tnote{4} & 0.06/0.02/0.04\tnote{4} & - & No fine-tuning \\
 
Wang et al.\cite{wang_sole_2023} & INT8/FP32  & - & DeiT, Swin, BERT & 28 & 1 & - &- & - & No fine-tuning \\

Liu et al.\cite{liu_consmax_2024} & INT8-FP\tnote{5}  & - & GPT-2 & 16 & 1.25 & 800 & 0.2 & - & Training \\

\textbf{Our} & BF16 & 1.62e-9 & GPT-2, ViT & 12 & 1 & 968\tnote{6}  & 7.1\tnote{6} & 0.45\tnote{6} & No fine-tuning\\
\bottomrule
\end{tabular}
    \begin{tablenotes}
        \item[*] Results are reported only for standalone designs (all synthesis results except for \cite{kim_hardware-efficient_2024}). For our design, we present the frequency of the full cluster and the post-layout area.
        \item[†] Denotes peak throughput, which may differ from average throughput.
        \item[1] The precision of the design is adjustable. The first value corresponds to the lowest precision setting, while the second value represents the highest precision setting ($P=3$) evaluated in the referenced paper.
        \item[2] The accelerator supports two input precisions: FX8 (first) and FX16 (second). For FX16, the reported results correspond to the version with a parallelization factor of 8.
        \item[3] Internal computations are performed in fixed-point format, with input and output values converted from and to floating-point format.
        \item[4] Values are reported for INT32, FP16, and FP32, respectively.
        \item[5] Internal computations are performed in floating-point format, with input and output values converted from and to INT8.
        \item[6] For our design, the reported area corresponds to the EXP unit per core, while the power and throughput are averaged over the entire Softmax operation per core.
      \end{tablenotes}
\label{tab:softmax_comparison}
\end{threeparttable}
}
\end{table*}

%% file: TEXT/06_Comparison_with_SoA.tex
\section{Comparison with the State-of-the-Art}
\label{subsec:soa}

We compare our solution to state-of-the-art Softmax accelerators evaluated for Transformer models, as shown in \autoref{tab:softmax_comparison}. Unlike fully custom datapaths for all Softmax operations, our approach introduces an \gls{isa} extension to accelerate only the exponential function while optimizing the remaining operations in software. This hybrid method balances efficiency and flexibility, supporting a broader range of applications at a low cost.

Our approach employs BF16 precision and achieves a mean squared error (MSE) of $1.62\mathrm{e}{-9}$, which is comparable to fixed-point approximations by Zhu et al. \cite{zhu_efficient_2020} and Kim et al. \cite{kim_hardware-efficient_2024}. In addition to this MSE, we demonstrate that our approximation preserves FP32/BF16 accuracy of GPT-2 and ViT-B, as detailed in \autoref{subsec:accuracy}. Most other works do not evaluate their methods on \glspl{llm} but rather focus on smaller, encoder-only models. Although Liu et al. \cite{liu_consmax_2024} achieves convergence to the same perplexity as the original GPT-2 during training, it remains unclear whether this approach can be applied without fine-tuning. Moreover, their architecture is designed for INT8 inputs/outputs while internally utilizing FP16 precision.

Other works primarily report post-synthesis evaluations (except for \cite{kim_hardware-efficient_2024}), omitting factors such as clock tree implementation and physical design. They also exclude timing, area, and power overheads arising from the integration of the custom datapaths into complex compute systems, making a thoroughly fair comparison impractical. Furthermore, while we report the average throughput per core over the entire Softmax computation, Zhu et al. \cite{zhu_efficient_2020} and Kim et al. \cite{kim_hardware-efficient_2024} provide only peak throughput values, which neglect iterations required for sequence lengths exceeding the hardware-supported size of 8, as well as memory operations. Despite these limitations, our hybrid hardware-software approach, with a compact area footprint of \qty[detect-all=true]{968}{\micro\meter^2} per core, achieves 1.1$\times$ better area efficiency (in terms of \qty[detect-all=true]{}{Op \per cycle \per \milli\meter^2}) compared to the high-precision version of \cite{zhu_efficient_2020} and only 1.7$\times$ lower area efficiency than the low-precision version, without compromising flexibility. Notably, our approach does not require fine-tuning or quantization. Furthermore, our method delivers 1.4$\times$ greater area efficiency than the FX16 version of \cite{kim_hardware-efficient_2024} while having 2.4$\times$ lower area efficiency compared to the FX8 version. The lower power efficiency compared to \cite{kim_hardware-efficient_2024} stems from the focus on optimizing the exponential function with higher precision, whereas \cite{kim_hardware-efficient_2024} employs a softmax-specific hardware implementation with reduced fixed-point precision. Additionally, the reported power consumption accounts for the entire core over the full softmax computation, rather than only the exponential unit, with power consumption being 3.2$\times$ (FX8) and 7.4$\times$ (FX16) lower than that of \cite{kim_hardware-efficient_2024}.


%% file: TEXT/07_Conclusion.tex
\section{Conclusion}
This work proposes a novel method to accelerate the Softmax function, a key bottleneck in Transformer models, by integrating a custom exponential instruction into the RISC-V Snitch architecture. Through hardware/software co-design, the approach achieves up to 162.7$\times$ speedup, with 5.8$\times$ and 3.6$\times$ reductions in latency and energy for GPT-2 and ViT models. This research demonstrates the potential of RISC-V for energy-efficient AI in resource-constrained settings, balancing precision, power, and simplicity.